\documentclass[conference,a4paper,10pt]{IEEEtran}
\IEEEoverridecommandlockouts

\usepackage{amsmath,amssymb}
\usepackage{graphicx}
\usepackage{mathtools}

\usepackage{fancyhdr} 

\newcommand{\T}{\mathrm{T}}
\newcommand{\diag}{\operatorname{blkdiag}}
\newcommand{\matId}[1][]{\mathbf{I}_{#1}}
\newcommand{\matnul}[1][]{\mathbf{0}_{#1}}
\newcommand{\stav}[1][]{\mathcal{X}_{#1}}
\newcommand{\odh}[1][\lok]{\hat{\mathcal{X}}_{#1}}
\newcommand{\kov}[1][\lok]{\mathbf{P}_{#1}}
\newcommand{\mez}[1][]{\mathbf{\Pi}_{#1}}
\newcommand{\eloid}[1]{\varepsilon^{#1}}
\newcommand{\veku}[1][]{\mathbf{u}_{#1}}
\newcommand{\vekx}[1][]{\mathbf{x}_{#1}}
\newcommand{\matOm}[1][]{\Omega_{#1}}
\newcommand{\vahaW}[1][\lok]{\mathbf{W}_{#1}}
\newcommand{\lok}{i}
\newcommand{\slouc}{F}
\newcommand{\minf}[1][]{\mathbf{Y}_{#1}}
\newcommand{\matH}[1][\lok]{\mathbf{H}_{#1}}
\newcommand{\vahalam}{\lambda}
\newcommand{\vahami}{\mu}
\newcommand{\vahaom}[1][]{\omega_{#1}}
\newcommand{\imez}[1][]{\mathbf{C}_{#1}}
\newcommand{\idealni}{^{\star}}
\newcommand{\optimalni}{^{\ast}}
\newcommand{\dolni}[1][]{\mathbf{O}_{#1}}
\newcommand{\vahaw}{w}
\newcommand{\ci}{^{\mathrm{CI}}}
\newcommand{\ici}{^{\mathrm{ICI}}}
\newcommand{\matX}{\mathbf{X}}
\newcommand{\mnozX}[1][]{\mathbb{X}_{#1}}
\newcommand{\dmez}{\mathbf{B}}
\newcommand{\matN}{\mathbf{N}}
\newcommand{\matD}{\mathrm{D}}
\newcommand{\sloz}{j}
\newcommand{\prvd}{d_{\sloz}}
\newcommand{\matV}{\mathrm{V}}
\newcommand{\matM}{\mathrm{M}}
\newcommand{\matE}[1][]{\mathrm{E}_{#1}}
\newcommand{\matSig}[1][]{\Sigma_{#1}}
\newcommand{\uhel}{\varphi}
\newcommand{\veke}{\mathrm{e}}
\newcommand{\barmez}[1][]{\bar{\mathbf{\Pi}}_{#1}}

\begin{document}
\title{Dual Approach to Inverse Covariance Intersection Fusion
\thanks{The work was supported by the Ministry of Education, Youth and Sports of 
the Czech Republic under project ROBOPROX - Robotics and Advanced 
Industrial Production
CZ.02.01.01/00/22\_008/0004590.}
}

\author{\IEEEauthorblockN{Ji\v{r}\'{i}~Ajgl\quad Ond\v{r}ej~Straka}
\IEEEauthorblockA{\textit{European Centre of Excellence -- New Technologies for Information Society and}\\ \textit{Department of Cybernetics,
Faculty of Applied Sciences, University of West Bohemia}\\
Pilsen, Czech Republic\\
jiriajgl@kky.zcu.cz, straka30@kky.zcu.cz}
}

\maketitle

\thispagestyle{fancy}
\addtolength{\headheight}{35pt}
\addtolength{\topmargin}{-15pt}
\addtolength{\headsep}{-20pt}
\renewcommand{\headrulewidth}{0pt}
\lhead{\copyright 2024 IEEE. Personal use of this material is permitted. Permission from IEEE must be obtained for all other uses, in any current or future media, including reprinting/republishing this material for advertising or promotional purposes, creating new collective works, for resale or redistribution to servers or lists, or reuse of any copyrighted component of this work in other works.
\
{https://ieeexplore.ieee.org/document/10705759}, doi:10.1109/MFI62651.2024.10705759
}
\cfoot{}

\begin{abstract}
 Linear fusion of estimates under the condition of no knowledge of correlation of estimation errors has reached maturity. On the other hand, various cases of partial knowledge are still active research areas. A frequent motivation is to deal with "common information" or "common noise", whatever it means. A fusion rule for a strict meaning of the former expression has already been elaborated. Despite the dual relationship, a strict meaning of the latter one has not been considered so far. The paper focuses on this area. The assumption of unknown "common noise" is formulated first, analysis of theoretical properties and illustrations follow. Although the results are disappointing from the perspective of a single upper bound of mean square error matrices, the partial knowledge demonstrates improvement over no knowledge in suboptimal cases and from the perspective of families of upper bounds.
\end{abstract}


\section{Introduction}
 The set of positive semi-definite matrices is a convex cone that defines a partial order, which is called Loewner order \cite{Horn:13}. That is, multiplication of a positive semi-definite matrix by a positive scalar produces a positive semi-definite matrix and the sum of two such matrices is again a positive semi-definite matrix. The partial order is defined by the relation "greater than or equal", which holds if the difference of the matrices is positive semi-definite. Two matrices need not be comparable and least upper bounds need not exist, i.e.\ many incomparable upper bounds can be minimal in the sense that there is no smaller upper bound.
 
 The matrices are related to ellipsoids \cite{Kurzhanski:97,Ajgl:23a}, which are generalisations of ellipses in multiple dimensions. The relation "greater than or equal" means that the second ellipsoid is included in the first one (lies inside). Incomparable ellipsoids are such ellipsoids that neither one lies in the other. To observe the meaning on minimal upper bounds, see that there are many ellipses that tightly circumscribe a rectangle, but they are not comparable.
 
 Covariance matrices of estimation errors (or the mean square error matrices) describe the uncertainty in many estimation problems \cite{Kailath:00,Simon:06}. It may be critical to not undervalue the uncertainty, e.g.\ in collision avoidance, even if specific processing of data is applied. Linear fusion \cite{Chang:97}--\nocite{Li:03}\cite{Chong:17} combines estimates, while knowledge of correlations of the estimation errors is a prerequisite for optimality.
 
 Fusion under unknown correlations has been studied for decades \cite{Julier:97}--\nocite{Chen:02a,Reinhardt:15}\cite{Forsling:22}. Unfortunately, the notation in the early works is confused and breeds erroneous interpretations. Most successors neglect the essence of the proof of "conservativeness", but a meticulous reader should be able infer the required definitions and assumptions. The name of Covariance Intersection (CI) for the fusion rule stems from the necessary condition on the bound-optimal covariance matrix of estimation error of the fused estimate. The optimisation is related to bounding admissible matrices by upper bounds, i.e.\ to circumscribing ellipsoids around unions of admissible fused ellipsoids.
 
 Partial knowledge of correlation restricts the sets of admissible matrices. That is, the resulting bound-optimal matrices cannot be worse (i.e.\ larger) than those for the case of no knowledge. A correct methodology is to parametrise the sets explicitly; an indirect description of the sets is misapplication-prone. Some cases of partial knowledge are well-motivated, for example \cite{Julier:01}--\nocite{Uhlmann:03}\cite{Wu:18}, where assumptions are widely understood and their validity can be maintained in dynamical systems. This paper stems from the partial knowledge treated in \cite{Noack:17}--\nocite{Noack:17a}\cite{Orguner:17}, where the name of Inverse Covariance Intersection (ICI) for the fusion rule is motivated by rather complicated considerations. The underlying assumption of ICI is existence of "common information", that is however not known.

 Theoretical properties of ICI have been studied extensively by the authors. The assumption has been formulated mathematically in terms of the sets of admissible matrices in \cite{Ajgl:19}. The set of the original upper bounds has been extended there, which unveiled that the optimal fusion rule is not unique. A proof that the new set contains strictly better bounds has been provided in \cite{Ajgl:18c} and the existence of non-zero lower bounds has been discovered. Inconvenient differences from CI have been pointed out in \cite{Ajgl:18b}. Namely, the fusion rule does not guarantee its properties, if the input matrices are replaced by arbitrary upper bounds, if the estimates are reused and if the rule is used to combine estimates in dynamical systems. The possibility to process multiple estimates sequentially has been reported as well, but the proof is invalid there and its correct version was published later. An explicit description of the set of admissible matrices in the terms of union of ellipsoids has been treated in \cite{Ajgl:19b}. The union has to be computed numerically and the derivation proves optimality of the set of bounds, i.e.\ that the union is tightly circumscribed by the set of upper-bounding ellipses, although the individual ellipses need not be tight. Extension to fusion of partial estimates is elaborated in \cite{Ajgl:19c}. The weak and strong versions of the assumption are proposed with the conclusion that the CI rule can be improved only in the case of the strong assumption for the combination a partial estimate with a full-state estimate. Last, generalisation to multiple full-state estimates has been presented in \cite{Ajgl:20a}, where the assumption is stated first and a suboptimal family of upper bounds is parametrised subsequently.
 
 The concepts of "common information" or "data incest" are frequently used in the literature, but according to the authors' knowledge, they are nowhere defined properly. Although the intuition for static problems with independent measurement errors is clear, it fails in more general settings. The dual concept "common process noise" is clear in the qualitative sense, but different multiples of the same noise in different estimates blur the quantitative point of view. Since neither "common information" nor "common noise" cause symmetric correlation in dynamical systems, they are rather slang words than technical terms. Their duality nevertheless raises the question whether there is a dual approach to ICI with the "common information" formulated in a very strict sense.
 
 The goal of the paper is to prospect the area, i.e.\ to define a strict sense of unknown "common noise", to design upper bounds and to investigate the fusion.

 The paper is organised as follows. Section~\ref{sec:formulace} provides mathematical formulations and proposes a new family. Section~\ref{sec:analyza} analyses the bounds and fusion; duality with ICI is quoted there. Section~\ref{sec:priklady} provides illustration and a comparison with CI. The contribution is summarised in Section~\ref{sec:zaver}.

\section{Problem Formulation}
 \label{sec:formulace}
 A new assumption of partial knowledge of correlation is formulated here. First, the technical background is treated in Section~\ref{ssec:castecna_znalost}. Inverse Covariance Intersection, i.e.\ the motivation of the paper, is reminded next in Section~\ref{ssec:inverzni_prunik}. The new dual assumption is finally proposed in Section~\ref{ssec:nova_trida}.
 
 \subsection{Partial Knowledge of Correlation}
  \label{ssec:castecna_znalost}
  Let $\kov$ with $\lok=1,2$ be two positive definite matrices and $\kov[]$ be an element of a family of positive semi-definite block matrices given by 
  \begin{equation}
   \kov[]=\begin{bmatrix}\kov[1] & \kov[12]\\ \kov[12]^\T & \kov[2]\end{bmatrix},
  \end{equation}
  where the block parameter $\kov[12]$ guarantees positive semi-definiteness of the Schur complement $\kov[2]-\kov[12]^\T \kov[1]^{-1} \kov[12]$.
  
  Matrix upper bounds $\mez$ of the family of $\kov[]$ are considered in the Loewner sense, i.e.\ that the difference $\mez-\kov[]$ is positive semi-definite for each admissible $\kov[12]$. Construction of $\mez$ for specific families of $\kov[12]$ is the topic of this paper.

  The motivation is the fusion of estimates under unknown correlation, where the construction is the crucial technical step. The reader can skip the following paragraph, which contains standard fusion formulas unnecessary for the construction of $\mez$; see e.g.\ \cite{Reinhardt:15}, \cite{Forsling:22}, \cite{Ajgl:18c}.
  
  For a variable $\stav$ and given matrices $\matH$, $\lok=1,2$, let estimates $\odh$ of $\matH \stav$ be available, together with corresponding mean square error matrices $\kov$. Linear fusion designs an estimate $\odh[\slouc]$ by the product $\vahaW[] \odh[]$ with $\vahaW[]=[\vahaW[1],\vahaW[2]]$, $\odh[]=[\odh[1]^\T,\odh[2]^\T]^\T$ and a regularity condition $\vahaW[] \matH[]=\matId$ with $\matH[]=[\matH[1]^\T,\matH[2]^\T]^\T$ and $\matId$ denoting the identity matrix. The mean square error matrix $\kov[\slouc]$ of the fused estimate $\odh[\slouc]$ is then given by $\kov[\slouc]=\vahaW[] \kov[] \vahaW[]^\T$. For known $\kov[12]$, the optimal weight and matrix are given by $\vahaW[]\idealni=(\matH[]^\T \kov[]^{-1} \matH[])^{-1} \matH[]^\T \kov[]^{-1}$ and $\kov[\slouc]\idealni=\vahaW[]\idealni \kov[] (\vahaW[]\idealni)^\T=(\matH[]^\T \kov[]^{-1} \matH[])^{-1}$ for any standard criterion such as trace or determinant. For unknown $\kov[12]$, a family of upper bounds $\mez$ of $\kov[]$ is constructed first. The weight $\vahaW[]\optimalni$ is constructed analogously by $\vahaW[]\optimalni=(\matH[]^\T \mez^{-1} \matH[])^{-1} \matH[]^\T \mez^{-1}$, where a single $\mez$ is searched according to an optimality criterion. Multiple bounds of the mean square error of $\vahaW[]\optimalni \odh[]$ can be found in the form $\mez[\slouc]\optimalni=\vahaW[]\optimalni \barmez (\vahaW[]\optimalni)^\T$, where the bound $\barmez$ can differ from $\mez$ used in the construction of the weight $\vahaW[]\optimalni$; the case $\barmez=\mez$ produces $\mez[\slouc]\optimalni=(\matH[]^\T \mez^{-1} \matH[])^{-1}$. Geometrically, the combination $\vahaW[] \odh[]$ corresponds to a mapping to a lower-dimensional space \cite{Ajgl:23a}. Selected existing families of $\kov[12]$ are discussed next.
 
 \subsection{Existing Fusion Rules}
  \label{ssec:inverzni_prunik}
  The family of $\kov[12]$ that is prospected in \cite{Noack:17}, \cite{Ajgl:19c}, \cite{Ajgl:20a} in the context of ICI fusion is motivated by a specific structure of estimation errors. Informally, it assumes that a common information is present in both estimates and that individual measurements are independent. For full-state estimates given by $\matH=\matId$, the idea can be expressed explicitly in the terms of $\kov[12]$ as
  \begin{equation}
   \kov[12]\ici \in \{\kov[1] \minf \kov[2]| \minf=\minf^\T, \minf \ge \matnul, \kov^{-1} \ge \minf, \lok=1,2\}. \label{eq:ici_korelace}
  \end{equation}
  
  The ICI fusion rule is based on upper bounds $\mez$ of $\kov[]\ici$ that are given by \cite{Ajgl:19}, \cite{Ajgl:18c} as
  \begin{equation}
   \mez\ici = \begin{bmatrix}\kov[1] & \matnul \\ \matnul & \kov[2]\end{bmatrix} + \begin{bmatrix}\tfrac{1}{\sqrt{\vahami}}\kov[1]\\ \sqrt{\vahami}\kov[2]\end{bmatrix} \tfrac{1}{2}\imez\ici \begin{bmatrix}\tfrac{1}{\sqrt{\vahami}}\kov[1]\\ \sqrt{\vahami}\kov[2]\end{bmatrix}^\T,
  \end{equation}
  where the parameter $\vahami$ is positive and the matrix  $\imez\ici$ has the property that $\veku^\T (\imez\ici)^{-1} \veku \le \max\{\veku^\T \kov[1] \veku, \veku^\T \kov[2] \veku\}$ holds for all vectors $\veku$. For example, a family of $\imez\ici$ parametrised by $\vahaom$ fulfilling $0\le \vahaom \le 1$ is given by
  \begin{equation}
   \imez\ici =(\vahaom \kov[1] + (1-\vahaom) \kov[2])^{-1}.
  \end{equation}
  
  The generally-applicable bounds for unknown correlations, $\kov[12]\ci \in \{\kov[12]| \kov[] \ge \matnul\}$, considered in the CI fusion can be parametrised by positive $\vahalam$ as
  \begin{equation}
   \mez\ci= \begin{bmatrix}(1+\frac{1}{\vahalam}) \kov[1] & \matnul \\ \matnul & (1+\vahalam) \kov[2]  \end{bmatrix}. \label{eq:mezci}
  \end{equation}
  That is, the matrix $\mez\ci$ is a block-diagonal matrix that can be denoted as $\diag\big(\vahaw^{-1} \kov[1],(1-\vahaw)^{-1}\kov[2])$ for $\vahaw=(1+\tfrac{1}{\vahalam})^{-1}$. 
    
 \subsection{Proposed Family of Correlation}
  \label{ssec:nova_trida}
  The focal idea of this paper is to investigate the formulation dual to the common information. Informally, it is assumed that each estimate contains a sum of a common noise and an individual noise, where the individual noises are independent.  The assumption is similarly academic as the former one \eqref{eq:ici_korelace}, since estimation errors can seldom be decomposed purely into noise-terms or into information-terms. Nevertheless, the hope is that the equations will contribute to the mathematics of the fusion under unknown (or partially known) correlation and that the paper will contribute to its philosophy.
  
  The proposed family of $\kov[12]$ for full-state estimates, $\matH=\matId$, is given by
  \begin{equation}
   \kov[12] \in \{\matX| \matX=\matX^\T, \matX \ge \matnul, \kov \ge \matX, \lok=1,2\}. \label{eq:korelace}
  \end{equation}
  
  To observe that such $\kov[12]$ define valid joint matrices $\kov[]$, a decomposition to two positive semi-definite matrices can be performed,
  \begin{equation}
   \kov[]=\begin{bmatrix}\kov[1]-\matX & \matnul \\ \matnul & \kov[2]-\matX \end{bmatrix} + \begin{bmatrix}\matId \\ \matId\end{bmatrix} \matX \begin{bmatrix}\matId \\ \matId\end{bmatrix}^\T.
  \end{equation}
  Finally, note that an extension to multiple full-state estimates would be analogous to \cite{Ajgl:20a}, i.e.\ one common $\matX$ for all cross-correlations and individual $\kov-\matX$ on the diagonal.

\section{Analysis}
 \label{sec:analyza}
 Analysis of the problem for the proposed assumption follows. Upper bounds of the proposed family are designed and discussed in Section~\ref{ssec:horni_meze}. Section~\ref{ssec:dolni_mez} inspects a lower bound. Section~\ref{ssec:fuze} applies both bounds in linear fusion.

 \subsection{Upper Bounds}
  \label{ssec:horni_meze}
  The assumptions \eqref{eq:ici_korelace} and \eqref{eq:korelace} have similar form. If the first is adopted and the matrix $\kov[]$ is pre-multiplied and post-multiplied by a block-diagonal matrix with the blocks $\kov$, i.e.\ by $\diag(\kov[1],\kov[2])$, the second form is almost obtained. The difference is notational; the meaning of the terms $\kov$ and $\kov^{-1}$ is switched. Straightforward manipulations lead to the bounds 
  \begin{equation}
   \mez = \begin{bmatrix}\kov[1] & \matnul \\ \matnul & \kov[2]\end{bmatrix} + \begin{bmatrix}\tfrac{1}{\sqrt{\vahami}}\matId\\ \sqrt{\vahami}\matId\end{bmatrix} \tfrac{1}{2}\dmez \begin{bmatrix}\tfrac{1}{\sqrt{\vahami}}\matId\\ \sqrt{\vahami}\matId\end{bmatrix}^\T, \label{eq:mez}
  \end{equation}
  where the parameter $\vahami$ is positive and the matrix $\dmez$ has the property that $\vekx^\T \dmez^{-1} \vekx \le \max\{\vekx^\T \kov[1]^{-1} \vekx, \vekx^\T \kov[2]^{-1} \vekx\}$ holds for all vectors $\vekx$. For example, a family of $\dmez$ parametrised by $\vahaom$, $0\le \vahaom \le 1$, is given by
  \begin{equation}
   \dmez =(\vahaom \kov[1]^{-1} + (1-\vahaom) \kov[2]^{-1})^{-1}. \label{eq:dmez}
  \end{equation}
  It should be noted that this matrix is nominally equal to the fused CI bound for $\vahaom=\vahaw$, but is not related to the mean square error matrix of any fused estimate, since no fused estimate has been constructed so far.
  
  The proof of upper-bounding property is analogous to \cite{Ajgl:19} \cite{Ajgl:20a}. Positive semi-definiteness of the difference $\mez-\kov[]$ is to be verified,
  \begin{equation}
   \mez-\kov[]=\begin{bmatrix}\tfrac{1}{2\vahami} \dmez & \tfrac{1}{2} \dmez - \matX \\ \tfrac{1}{2} \dmez - \matX & \tfrac{\vahami}{2}\dmez\end{bmatrix}.
  \end{equation}
  The first block $\tfrac{1}{2\vahami} \dmez$ is positive definite by construction. The Schur complement $\tfrac{\vahami}{2}\dmez- (\tfrac{1}{2} \dmez - \matX) (\tfrac{1}{2\vahami} \dmez)^{-1} (\tfrac{1}{2} \dmez - \matX)$ is equal to $2 \vahami (\matX - \matX \dmez^{-1} \matX)$. Due to the construction of $\dmez$ and the assumption $\kov \ge \matX$, it holds $\dmez \ge \matX$ (see the "necessary CI condition" \cite{Chen:02a}). That is, the complement is positive semi-definite, q.e.d.
  
  The proposed bounds $\mez$ \eqref{eq:mez} are better than the general bounds $\mez\ci$ \eqref{eq:mezci}. The verification is analogous to \cite{Ajgl:18c}, \cite{Ajgl:20a}. For $\vahaom=\tfrac{1}{2}$ and $\vahami=\vahalam$, it holds that $\mez\ci-\mez$ is equal to 
  \begin{equation}
   \begin{bmatrix}\tfrac{1}{\vahalam}\big(\kov[1]-(\kov[1]^{-1}+\kov[2]^{-1})^{-1}\big) & -(\kov[1]^{-1}+\kov[2]^{-1})^{-1} \\ -(\kov[1]^{-1}+\kov[2]^{-1})^{-1} & \vahalam \big(\kov[2]-(\kov[1]^{-1}+\kov[2]^{-1})^{-1}\big)\end{bmatrix}.
  \end{equation}
  The application of standard equivalents of $(\kov[1]^{-1}+\kov[2]^{-1})^{-1}$, i.e.\ of the terms $\kov[1]-\kov[1](\kov[1]+\kov[2])^{-1}\kov[1]$, $\kov[1](\kov[1]+\kov[2])^{-1}\kov[2]$, $\kov[2](\kov[1]+\kov[2])^{-1}\kov[1]$ and $\kov[2]-\kov[2](\kov[1]+\kov[2])^{-1}\kov[2]$ gives
  \begin{equation}
   \mez\ci-\mez= \begin{bmatrix}\tfrac{1}{\sqrt{\vahalam}} \kov[1] \\ -\sqrt{\vahalam} \kov[2]\end{bmatrix} (\kov[1]+\kov[2])^{-1} \begin{bmatrix}\tfrac{1}{\sqrt{\vahalam}} \kov[1] \\ -\sqrt{\vahalam} \kov[2]\end{bmatrix}^\T, \label{eq:mezci_mez}
  \end{equation}
  which is a positive semi-definite matrix, q.e.d.
  
  Contrary to ICI \cite{Ajgl:18b}, the proposed matrices $\mez$ remain to bound $\kov[]$ even if matrices $\kov$ are replaced by greater matrices during the construction \eqref{eq:mez}--\eqref{eq:dmez}. The reason is that $\mez$ is increasing in $\kov$. From another perspective, see also that inflating the matrices $\kov$ leads to a set that is larger than the one in \eqref{eq:korelace}. The resulting upper bounds need not be tight, but they are guaranteed to be bounds. 
    
 \subsection{Lower Bound}
  \label{ssec:dolni_mez}
  The same reasoning as in the previous section leads to a lower bound of $\kov[]$, i.e.\ to such a matrix $\dolni$ that $\kov[] \ge \dolni$ holds for any $\kov[12]$ given by \eqref{eq:korelace}. The form is analogous to \cite{Ajgl:18c}, \cite{Ajgl:20a}; it holds
  \begin{equation}
   \dolni = \begin{bmatrix}\kov[1] \\ \kov[2]\end{bmatrix} (\kov[1]+\kov[2])^{-1} \begin{bmatrix}\kov[1] \\ \kov[2]\end{bmatrix}^\T.
  \end{equation}
  
  The proof of the lower-bounding property verifies positive semi-definiteness of the difference $\kov[]-\dolni$, which can be expressed by application of the above-used standard equivalents as  
  \begin{equation}
   \kov[]-\dolni=\begin{bmatrix} \matN & \matX - \matN \\ \matX - \matN & \matN \end{bmatrix},
  \end{equation}
  where $\matN$ is given by $\matN=(\kov[1]^{-1}+\kov[2]^{-1})^{-1}$ and is positive definite. The Schur complement $\matN - (\matX-\matN) \matN^{-1} (\matX-\matN)$ is equal to $2 \matX - \matX \matN^{-1} \matX$. Due to the assumption $\kov \ge \matX$, it holds $\matX^{-1} \ge \kov^{-1}$, which implies $2 \matX^{-1} \ge \matN^{-1}$. Thus, the complement is positive semi-definite, q.e.d.
  
 \subsection{Linear Fusion}
  \label{ssec:fuze}
  The assumption of common information \cite{Noack:17}, \cite{Ajgl:19} leads to optimal fusion weights $\vahaW[]$ different than in the case of unknown correlations. The optimal bounds $\mez[\slouc]$ of the fused estimates are incomparable on the philosophical level, since they correspond to different fused estimates, but the nominal values favour the partial knowledge. This section discovers that this is not the case of the common noise assumption.
  
  The fusion weights $\vahaW[]\ci$ derived from the $\mez\ci$ bounds \eqref{eq:mezci} are given by 
  \begin{equation}
   \vahaW[]\ci= \big(\vahaw \kov[1]^{-1} + (1-\vahaw) \kov[2]^{-1}\big)^{-1} \begin{bmatrix}\vahaw \kov[1]^{-1} & (1-\vahaw) \kov[2]^{-1} \end{bmatrix}. \label{eq:vahaWci}
  \end{equation}
  Since the new bounds $\mez$ are better than the $\mez\ci$ bounds for $\vahami=\vahalam$ and $\vahaw=(1+\tfrac{1}{\vahalam})^{-1}$, $\mez\ci \ge \mez$, a meaningful comparison of the fused bounds for the same fusion weight $\vahaW[]\ci$ has to produce the inequality $\vahaW[]\ci \mez\ci \vahaW[]\ci \ge \vahaW[]\ci \mez \vahaW[]\ci$. Nevertheless, degeneration to equality occurs. To see validity of $\vahaW[]\ci (\mez\ci - \mez) (\vahaW[]\ci)^\T=\matnul$, combine \eqref{eq:vahaWci} with \eqref{eq:mezci_mez}. In particular, it holds $[\vahaw \kov[1]^{-1}, (1-\vahaw) \kov[2]^{-1}] [\tfrac{1}{\sqrt{\vahalam}} \kov[1], -\sqrt{\vahalam} \kov[2]]^\T=$\\ $(\vahaw\tfrac{1}{\sqrt{\vahalam}} - (1-\vahaw) \sqrt{\vahalam}) \matId$ and the substitution of the parameters gives $\tfrac{\vahalam}{\vahalam+1}\tfrac{1}{\sqrt{\vahalam}} - \tfrac{1}{\vahalam+1} \sqrt{\vahalam}=0$.
  
  The above comparison focuses on a single bound. For the optimal weight, the partial knowledge of correlation \eqref{eq:korelace} does improve the fusion in the sense of whole families, but the optimal fused bound is not improved. For suboptimal weights, the fused bound can be improved. Further insights require explicit parametrisation of some admissible matrices $\matX$ in Section~\ref{ssec:parametry}. Graphical illustration is given in Section~\ref{ssec:elipsy}.
  
  Next, it will be shown that the weights $\vahaW[]\ci$ are optimal with respect to the specific bounds in the new family \eqref{eq:mez}. For $\matH[]=[\matId, \matId]^\T$, an application of matrix inversion lemma expresses the product $\matH[]^\T \mez^{-1}$ as 
  \begin{align}
   & \matH[]^\T \mez^{-1} = \begin{bmatrix}\kov[1]^{-1} & \kov[2]^{-1} \end{bmatrix} - (\tfrac{1}{\sqrt{\vahami}} \kov[1]^{-1} + \sqrt{\vahami} \kov[2]^{-1})  \times \nonumber \\
   & \ \times (2\dmez^{-1} + \tfrac{1}{\vahami} \kov[1]^{-1} + \vahami \kov[2]^{-1})^{-1} \begin{bmatrix}\tfrac{1}{\sqrt{\vahami}} \kov[1]^{-1} & \sqrt{\vahami} \kov[2]^{-1}\end{bmatrix}. \label{eq:matH_mez}
  \end{align}
  The matrix $\dmez$ designed by equation \eqref{eq:dmez} leads to the equality $2\dmez^{-1} + \tfrac{1}{\vahami} \kov[1]^{-1} + \vahami \kov[2]^{-1} = (2 \vahaom + \tfrac{1}{\vahami}) \kov[1]^{-1} + \big(2(1-\vahaom) + \vahami\big) \kov[2]^{-1}$. If it holds $\vahaom=\tfrac{1}{2}$, the ratios of the multipliers of the inverse matrices $\kov^{-1}$ in the parentheses in \eqref{eq:matH_mez} are thus given by
  \begin{equation}
   \frac{\frac{1}{\sqrt{\vahami}}}{2 \vahaom + \frac{1}{\vahami}} = \frac{\sqrt{\vahami}}{1+\vahami}, \qquad \frac{\sqrt{\vahami}}{2(1-\vahaom) + \vahami} = \frac{\sqrt{\vahami}}{1+\vahami}.
  \end{equation}
  Equality of these ratios simplifies the product \eqref{eq:matH_mez} to
  \begin{equation}
   \matH[]^\T \mez^{-1} = \begin{bmatrix}(1-\frac{1}{1+\vahami})\kov[1]^{-1} & (1-\frac{\vahami}{1+\vahami})\kov[2]^{-1} \end{bmatrix} 
  \end{equation}
  and the identity with \eqref{eq:vahaWci} for $\vahami=\vahalam$ and $\vahaw=(1+\tfrac{1}{\vahalam})^{-1}$ is evident.
  
  Last, note that the new bounds $\mez$ offer a non-trivial fused lower bound $\dolni[\slouc]$ for a fixed fusion weight $\vahaW[]$, i.e.\ for $\dolni[\slouc]= \vahaW[] \dolni \vahaW[]^\T$, it holds $\kov[\slouc] \ge \dolni[\slouc]$ and $\dolni[\slouc] \ge \matnul$.

\section{Illustration}
 \label{sec:priklady}
 This section visualises the theory. Subsets of admissible parameters are presented in Section~\ref{ssec:parametry}. Ellipses are compared in Section~\ref{ssec:elipsy}.

 \subsection{Admissible parameters}
  \label{ssec:parametry}
  Graphical verification of upper-bounding property calls for carefully selected elements of the families of the parameters $\kov[12]$. Two subsets are proposed. 
  
  An easy way is to consider all maximal rank-$1$ matrices $\matX$ in \eqref{eq:korelace}. The subset $\mnozX[\vekx]$ is parametrised by unit vectors $\vekx$ as
  \begin{equation}
   \mnozX[\vekx]= \big\{\vekx \big(\max\{\vekx^\T \kov[1]^{-1} \vekx, \vekx^\T \kov[2]^{-1} \vekx\}\big)^{-1} \vekx^\T \big| \vekx^\T \vekx=1 \big\}. \label{eq:mnozX_vekx}
  \end{equation}
  
  The proposed alternative is to find maximal positive definite matrices $\matX$ such that $\kov[1] \ge \matX$ and $\kov[2] \ge \matX$. The parametrisation by $\matOm$ is discussed in \cite{Ajgl:17a}. First, the generalised eigenvalue decomposition of the pair $(\kov[1],\kov[2])$ is performed, i.e.\ if the columns of a matrix $\matV$ are the generalised eigenvectors and $\matD$ is a diagonal matrix with the corresponding generalised eigenvalues $\prvd$, it holds $\kov[1] \matV= \kov[2] \matV \matD$. The matrices $\kov[1]$, $\kov[2]$ are thus decomposed as $\kov[1]=\matV^{-\T} \matD \matV^{-1}$, $\kov[2]=\matV^{-\T} \matId \matV^{-1}$. Next, a matrix $\matM$ is introduced as $\matM=\max\{\matD,\matId\}$ with the maximum being taken over the pairs $(\prvd,1)$ of diagonal elements. The inverses of $\matD$ and $\matId$ are further expressed as
  \begin{align}
   \matD^{-1}&=\matM^{-1} + \matE[1]^\T \matSig[1]^{-1} \matE[1],\\
   \matId^{-1}&=\matM^{-1} + \matE[2]^\T \matSig[2]^{-1} \matE[2],
  \end{align}
  where $\matE[1]$ is composed of the $\sloz$-th rows of the identity matrix corresponding to $\prvd<1$ and $\matSig[1]$ is a diagonal matrix with the elements $(\tfrac{1}{\prvd}-1)^{-1}$ for these $\prvd$; the matrix $\matE[2]$ is composed analogously for $\prvd>1$ and $\matSig[2]$ contains $(1-\tfrac{1}{\prvd})^{-1}$ on its diagonal. Later, some square roots $\matSig[1]^{\frac{1}{2}}$ and $\matSig[2]^{\frac{1}{2}}$ of the corresponding matrices are computed, e.g.\ the Cholesky factors, and a matrix $\matE$ and a parametric matrix $\matSig[\matOm]$ are constructed
  \begin{equation}
   \matE=\begin{bmatrix}\matE[1] \\ \matE[2]\end{bmatrix}, \quad \matSig[\matOm]=\begin{bmatrix}\matSig[1] & \matSig[1]^{\frac{1}{2}} \matOm \matSig[2]^{\frac{\T}{2}} \\ \matSig[2]^{\frac{1}{2}} \matOm^\T \matSig[1]^{\frac{\T}{2}} & \matSig[2]\end{bmatrix}
  \end{equation}
  A family of lower bounds of $\kov[1]$, $\kov[2]$ is then given by
  \begin{equation}
   \mnozX[\matOm]= \{\matV^{-\T} (\matM^{-1} +  \matE^\T \matSig[\matOm]^{-1} \matE )^{-1} \matV^{-1} | \matId - \matOm \matOm^\T \ge \matnul\}. \label{eq:mnozX_matOm}
  \end{equation}
 
 \subsection{Ellipses}
  \label{ssec:elipsy}
  Symmetric positive semi-definite matrices can be visualised as ellipsoids \cite{Kurzhanski:97,Ajgl:23a}. For a matrix $\kov[]$, the boundary is given by $\eloid{\kov[]}=\{\veke|\veke^\T \kov[]^{-1} \veke=1\}$. In two-dimensional case, ellipses are obtained.
  
  Let the matrices $\kov$ be given by  
  \begin{equation}
   \kov[1]= \begin{bmatrix}9 & 3\\3 & 4\end{bmatrix}, \qquad \kov[2]=\begin{bmatrix}4 & -3\\-3 & 9\end{bmatrix}. \label{eq:priklad}
  \end{equation}
  The elements of \eqref{eq:mnozX_vekx} given by $\vekx=[\cos(\uhel),\sin(\uhel)]^\T$ with $\uhel\in\{-\frac{\pi}{4},0,\frac{\pi}{4},\frac{\pi}{2}\}$ will be used in the figures. The elements of \eqref{eq:mnozX_matOm} with $\matOm\in\{-1,-0.5,0,0.5,1\}$ will be used as well.
  
\begin{figure}
   \includegraphics[scale=0.8]{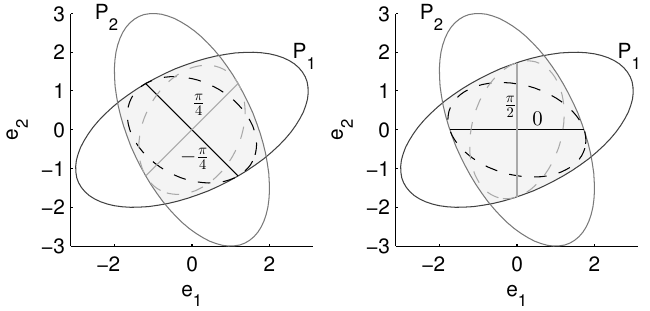}
   \caption{Individual ellipses $\eloid{\kov[1]}$ and $\eloid{\kov[2]}$ (solid) with four examples \eqref{eq:mnozX_vekx} of common noise ellipses $\eloid{\matX}$ (degenerated to line segments) and the corresponding ideal fusions $\eloid{\kov[\slouc]\idealni}$ (dashed ellipses).} 
   \label{fig:dolniX}
  \end{figure}

\begin{figure}
   \includegraphics[scale=0.8]{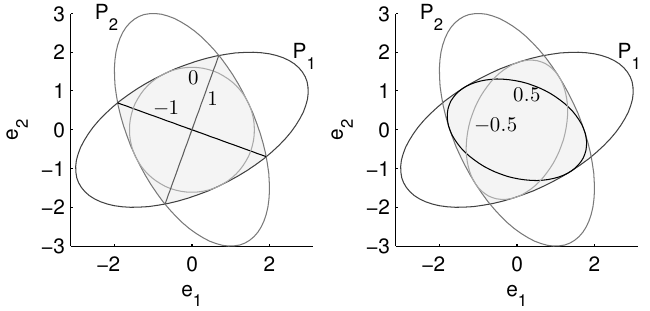}
   \caption{Individual ellipses $\eloid{\kov[1]}$ and $\eloid{\kov[2]}$ (solid) with five examples \eqref{eq:mnozX_matOm} of common noise ellipses $\eloid{\matX}$ (two degenerated to line segments), which are identical to the corresponding ideal fusions $\eloid{\kov[\slouc]\idealni}$.} 
   \label{fig:dolniOm}
  \end{figure}

  Figs.~\ref{fig:dolniX} and~\ref{fig:dolniOm} compare the individual ellipses $\eloid{\kov[1]}$ and $\eloid{\kov[2]}$ with the ellipses $\eloid{\matX}$ of admissible common noise. According to the problem formulation, the common noise ellipse lies inside the intersection of the individual ellipses. In the case \eqref{eq:mnozX_vekx} of rank-$1$ matrices $\matX$, the ellipse $\eloid{\matX}$ degenerates to a line segment. In the case \eqref{eq:mnozX_matOm} of matrix parameter $\matOm$, the ellipse $\eloid{\matX}$ degenerates in the case of reduced rank of $\matId-\matOm\matOm^\T$, i.e.\ for $\matOm=-1$ and $\matOm=1$ in this case \eqref{eq:priklad}. Each family $\mnozX[\vekx]$ and $\mnozX[\matOm]$ covers the intersection of $\eloid{\kov[1]}$ and $\eloid{\kov[2]}$ fully.
  
  Furthermore, Figs.~\ref{fig:dolniX} and~\ref{fig:dolniOm} show the ideal fusions for the particular matrices $\matX$. Since the noise is common to both estimation errors, the ideal ellipses $\eloid{\kov[\slouc]\idealni}$ contain the ellipses $\eloid{\matX}$; the dashed ellipses are strictly larger than the line segments in Fig.~\ref{fig:dolniX} (but they are not tightly inscribed into the intersection of the local ellipses) and the ellipses $\eloid{\kov[\slouc]\idealni}$ coincide with $\eloid{\matX}$ in Fig.~\ref{fig:dolniOm} (and they are inscribed tightly). In both the cases, the family of admissible ideal matrices $\kov[\slouc]\idealni$ covers the intersection of $\eloid{\kov[1]}$ and $\eloid{\kov[2]}$ fully. The figures demonstrate that no fusion can outperform the CI fusion in the terms of single upper bound \cite{Chen:02a}. Nevertheless, the partial knowledge \eqref{eq:korelace} of correlation has its value. These aspects are considered in the following figures.
  
  Fig.~\ref{fig:idealni} offers an inappropriate (but very traditional) comparison of ellipses $\eloid{\mez[\slouc]\optimalni}$. The ellipses correspond to fusions with different weights $\vahaW[]\optimalni$, i.e.\ to different random variables $\odh[\slouc]$, and they should not be compared directly. They can serve as a tool for designing the optimal fusion weight or as an illustration of a necessary condition. The top left figure shows the bounds $\eloid{\mez[\slouc]\optimalni}$ for $\mez$ given by \eqref{eq:mez} with $\dmez$ given by $\vahaom=\tfrac{1}{2}$ in \eqref{eq:dmez} and the parameter $\vahami$ given by $\vahami\in\{\tfrac{1}{3},1,3\}$. These bounds are identical to the CI bounds. The top right and bottom right figures show the bounds $\eloid{\mez[\slouc]\optimalni}$ for $\mez$ given by same $\vahami$ as before and by $\vahaom=0$ and $\vahaom=1$, respectively. As before, the limit cases for $\vahami \rightarrow 0$ or $\vahami \rightarrow \infty$ are $\mez[\slouc]\overset{\vahami \rightarrow 0}{=}\kov[2]$ and $\mez[\slouc]\overset{\vahami \rightarrow \infty}{=}\kov[1]$, respectively, but the intermediate $\vahami$ do not produce ellipses that tightly circumscribe the intersection of the individual ellipses $\eloid{\kov[1]}$ and $\eloid{\kov[2]}$; see the detail in the bottom left figure.
  \begin{figure}
   \includegraphics[scale=0.8]{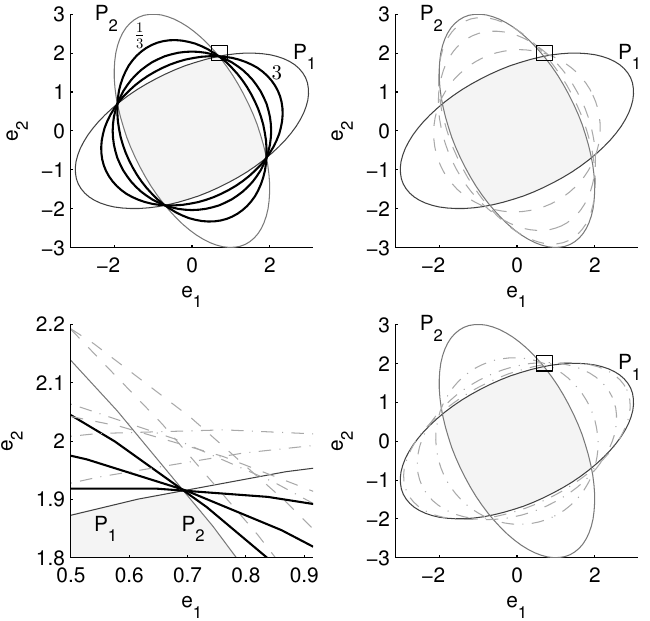}
   \caption{Forbidden comparison of individual ellipses $\eloid{\kov[1]}$ and $\eloid{\kov[2]}$ (solid) with examples of ideally fused upper bounds $\eloid{\mez[\slouc]\optimalni}$ for $\mez$ given by \eqref{eq:mez}--\eqref{eq:dmez} for three $\vahami$ in each figure and three $\vahaom$  in three respective figures. The bottom left figure compares the small rectangular area in detail.} 
   \label{fig:idealni}
  \end{figure}

\begin{figure}
   \includegraphics[scale=0.8]{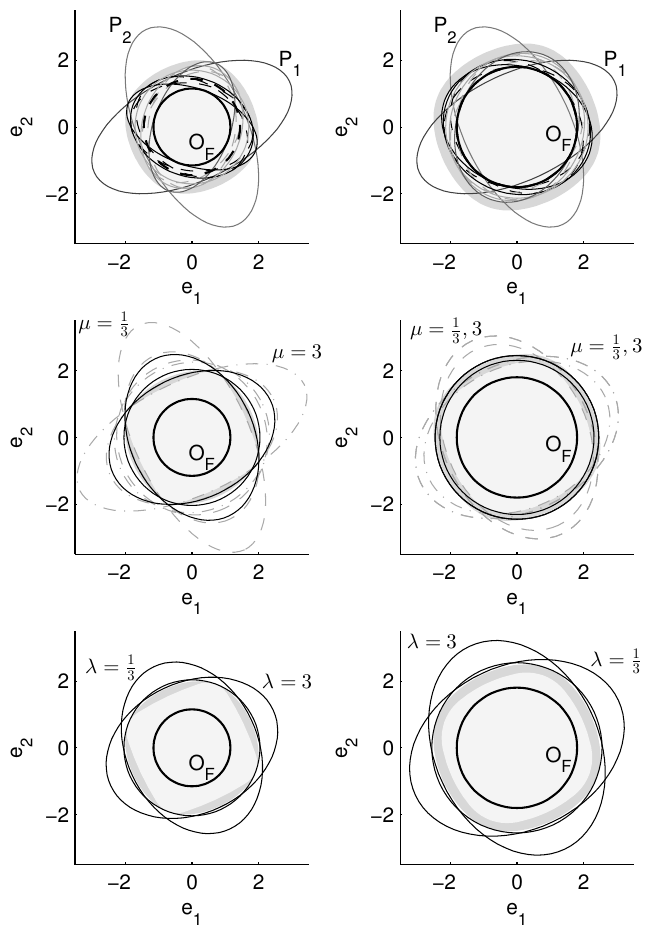}
   \caption{Fusion for two cases of fusion weight $\vahaW[]$ (left and right column, respectively). The top row shows $\eloid{\kov[\slouc]}$ for the examples of $\matX$ in Figs.~\ref{fig:dolniX} and \ref{fig:dolniOm} (in the light grey area) and the admissible area for all $\kov[12]$ (combined light and dark grey areas). The middle row shows examples of the proposed fused upper bounds, the bottom row shows examples of the existing bounds for all $\kov[12]$. Lower bounds are shown by thick ellipses (circles in this case).}
   \label{fig:fuze}
  \end{figure}

  Fig.~\ref{fig:fuze} considers the same parameters $\vekx$, $\matOm$, $\vahami$ and $\vahalam$ as before. Two different fusion weights $\vahaW[]$ are inspected. The left column considers $\vahaW[]\ci$ given by \eqref{eq:vahaWci} with $\vahaw=\tfrac{1}{2}$, the right column considers $\vahaW[]=[\tfrac{1}{2}\matId,\tfrac{1}{2}\matId]$. The union of all admissible ellipses $\eloid{\kov[\slouc]}$ for $\kov[12]$ given by \eqref{eq:korelace} is smaller than the union for all $\kov[12]\ci$; see the light and dark grey areas. The unions are tightly circumscribed by the ellipses for the corresponding families for fixed fusion weights and admissible matrices $\mez$ and $\mez\ci$; note however that the individual ellipses need not be tight. In the left column, families of $\vahaW[]\ci \mez (\vahaW[]\ci)^\T$ and $\vahaW[]\ci \mez\ci (\vahaW[]\ci)^\T$ are shown in the middle and bottom row, respectively. In the right column, the families $\vahaW[] \mez \vahaW[]^\T$ and $\vahaW[] \mez\ci \vahaW[]^\T$ are shown. For the choice $\vahami=\vahalam$ with $\vahaom=\frac{1}{2}$, the bold ellipses are the same in the left column, while in the right column, the new family produces smaller ellipses; see Section~\ref{ssec:fuze}. In the example \eqref{eq:priklad}, the bounds $\vahaW[] \mez (\vahaW[])^\T$ are the same for $\vahami=\frac{1}{3}$ as for $\vahami=3$ for any value of $\vahaom$. The figure demonstrates that the partial knowledge improves the CI fusion in the terms of  families of upper bounds. Also, note that there exist inscribed ellipses $\eloid{\dolni[\slouc]}$ of the admissible fused ellipses $\eloid{\kov[\slouc]}$. 


\section{Summary}
 \label{sec:zaver}
 Fusion under partial knowledge of correlation has been dealt with. A new family of matrices has been formulated first. The idea is dual to Inverse Covariance Intersection; instead of common information, a common identical noise is considered. The bounds and the fusion are analogous to the existing approach and generalisation to multiple estimates is straightforward. The joint bounds are better than the general bounds for no knowledge. Unfortunately, the fusion cannot improve the standard case in the sense of the best single bound, but it is superior in the sense of whole family of bounds. The illustrations explicitly consider the admissible correlations; the matrices have been visualised by ellipses and compared graphically last.


\end{document}